# The Java-Sumatra Aerial Mega-Tramway

Alexander A. Bolonkin[•]
C & R
1310 Avenue R, Suite 6-F
Brooklyn, New York 11229, USA
aBolonkin@juno.com
**Richard B. Cathcart**
Geographos
1300 West Olive Avenue, Suite M
Burbank, California 91506, USA
rbcathcart@charter.net

**ABSTRACT**

A mega-tramway based on the Indonesian islands of Sumatra and Java is proposed to span Sunda Strait. The Java-Sumatra Aerial Mega-Tramway (JSAMT) will be self-elevating and will regularly and cheaply launch passengers and cargoes via two conveyor belt-like facilities using standard winged shipping containers like those currently used by international trucking and ocean shipping industries that are volplaned across the Sunda Strait. The JSAMT will be a self-sustaining toll facility free of any requirement for international loans or funding guarantees for its construction. Its existence will remove any immediate need for an expensive to dig/maintain Nusantara Tunnel. We offer the formative basic technical specifications for the JSAMT and indicate some of the physical and cultural geographical facts underpinning our macro-engineering proposal; offshoots of a perfected and tested JSAMT may be installed at Palk Strait between India and Sri Lanka, the Gibraltar Strait and the Bering Strait by mid-21st Century.

## 1. INTRODUCTION

Sumatra is geographically defined by a 3 km-wide Malacca Strait and a 30 km-wide Sunda Strait; across the Sunda Strait is Java; further south Java is defined by the 18 km-wide Lomboc Strait. Approximately 245,000,000 people are citizens of the Republic of Indonesia; the nation's capital, Jakarta, situated on Java, may have ~21,000,000 residents by 2015. Jakarta Bay port facilities, which constitute a "mega-harbor"[1] are being improved at considerably expense.

Among the natural hazards affecting those persons living/working on Sumatra and Java, tsunamis, earthquakes, volcanoes and annual forest fires are most remarkable[2]. The 27 August 1883 Krakatoa hydro-volcanic explosion and tsunami in the Sunda Strait are famous as is the haze caused by forest fires on Sumatra. During 2002, six of the ten members of the Association of Southeast Asian Nations vowed to fight fire pollution (smoke) in the region[3]. Further development of comprehensive hazard mitigation is vitally necessary to reduce as much as possible the impact of natural and human-caused hazards. The region surrounding the Sunda Strait has a high potential to endure a possibly predictable volcanic eruption of Anak Krakatoa (volcanic ash cloud, air shockwaves, tsunami) and strong ground motion caused very strong earthquakes; in future, the Sunda Strait also may endure industrial hazards emanating from the Merak-Cilegon region and agricultural hazards in the Lampung and Ujung Kulon regions, where fallow land cleared for an anticipated season of crop planting, as well as natural forest felled and burned to create plantations, is subject to seasonal wildfires.

---

[•] Corresponding author.

For these reasons, the authors concur that a Nusantara Tunnel bored beneath the Sunda Strait's seafloor is an inappropriate macroproject for the Republic of Indonesia to undertake at this time[4]. As a viable alternative, we suggest the Java-Sumatra Aerial Mega-Tramway (JSAMT) based, in part, on the US Patent 6,494,143 awarded to Alexander Alexandrovich Bolonkin on 17 December 2002. If built, the JSAMT will be a truly remarkable 21st Century new kinetic aviation technology extension of the 260 m "Transporter Bridge" spanning the River Tyne since 1911 at Middlesbrough in England! Further, we conceive a cross-Palk Strait application of this particular technology will very likely be commenced due to the ongoing industrialization of the adjacent regions and construction of the Sethusamudram Ship Channel in a low-seismicity region[5].

## 2. GROUND AND AERIAL ISLAND LINKAGES

Indonesia's longest bridge, commenced during 2003 and scheduled for completion by 2008, is the cable-stayed Suramadu Bridge; it will connect the island of Java and the island of Madura with a modern high-speed four-lane highway. Soon, Indonesians will experience the "zoomscape"—the localized human experience of architecture and landscape that has been fundamentally altered by the globalization of speedy transportation technologies[6] — and come to appreciate their geographical surroundings in 21st Century synoptic apprehensions. The islands of Indonesia are connected by a network of hundreds of airports but the national highway and railway network remains fragmented; as a consequence, the logistics of intra-national commerce is complicated and, sometimes, uneconomical. Even common inter-modal standard shipping containers — like those stacked by the hundreds at Malaysia's seaports of Tanjung Pelepas and Port Klang — cannot yet be shifted throughout Indonesia with dispatch or economy[7]. In other words, the Republic of Indonesia's future industrialization requires a national policy encouraging the timely initial organization of a semi-fixed aerial linkage capable of moving containerized people and cargoes over the Sunda Strait at low cost; the Java-Sumatra Aerial Mega-Tramway is a logical, and affordable, kinetic aviation technology to accomplish that task within a reasonable period of time.

The Berlin Airlift (27 June 1948 to 12 May 1949) was an extraordinary first use of a new tool of policy, a system of transporting supplies by air when ground routes (highways, railways and canals) were blockaded[8]. Ultimately, about 2.1 million tonnes of supplies were ferried from western Germany to isolated Berlin in nearly 139,000 flights to Berlin through inbound routes ranging from 240 km to 450 km in length. (Weather forecasting for the two inbound routes and the single outbound air corridor was organized more thoroughly than any aviation operation previously devised.) What if an aerial mega-tramway connecting Berlin with a single site in western Germany had existed during 1948-49? Using calculations developed in **SECTION 4**, we have an answer for that alternate-history postulation: 2.1 millions metric ton of material could have been delivered in just 52500 standard containers of 40 tons each by two catapults (the first is located in Berlin and the second located in West Germany), 10 gliders (+2 in reserve, start weight of each glider is 240 tons) for 4×40=160 tonnes average capacity and flight frequency 0.1 hours (+20 minutes of loading, 16 minutes of flight, and +20 minutes of unloading), over a much shorter period of time 55 days instead of almost 320 days and 200 conventional transport aircraft! The Berlin Airlift advanced the logistics of airpower. We think the Java-Sumatra Aerial Mega-Tramway macroproject has the potential to truly revolutionize the Republic of Indonesia's industrial and commercial logistics for the suggested passengers and cargo transfer system after installation of universalized connection-disconnection devices atop the standard shipping containers[9]; JSAMT will consist of a 40-200 meter-long cable path with ground-based engines that propel standardized winged shipping containers 35 km across the 30 km-wide Sunda Strait. The distance across Palk Strait is about the same. Foreseeably, such winged containers will carry passengers, with each container having seats, baggage stowage and cabin crew. While the new Airbus

A380 will carry 555 persons, each winged container will likely safely carry about 100 persons so that loading/unloading logistics will be far, far simpler and efficient[10].

## 4. JAVA-SUMATRA AERIAL MEGA-TRAMWAY (JSAMT)

An aerial tramway is a type of aerial lift that uses comfortable cabins carrying passengers and pre-packaged cargoes; it constitutes the exact opposite of a water-traversing cable ferry that uses barges to haul passengers and cargoes across bodies of water even when water currents impose strong transverse flow. Naturally, suspended aerial tramway cabins can be violently jiggled by buffeting winds (head winds, trailing winds, tailwinds, cross winds). JSAMT will utilize kinetic aircraft to accelerate the standard weather-tight winged shipping containers to subsonic speed of 250 m/s until its speed decays to a safe landing speed of 50-60 m/s on a paved airdrome on the other side of Sunda Strait. An acceleration of 3g will not discomfort passengers with normal health and the paved airfield runway length actually necessary to bring the winged container to a halt is only about 1.5 km. The flight path spanning the Sunda Strait will still be subject to the vagaries of the weather as well as other naturally hazardous flight conditions (Fig. 1a and 1b). We anticipate a need for the internationally approved shipping containers—some of which will be loaded with passengers — to go no higher than 300 m - 500 m altitude above the Sunda Strait.

This system for Aerial Mega-Tramway—literally, a non-fixed aerial bridge—includes (Fig. 1) a closed-loop cable and drive station located on the Earth's surface. The cable is supported in the air by columns equipped with rollers. Each drive station has engines located on the ground and works on any cheap energy. The system works in the following way. The subsonic load glider (winged container, aircraft) starts from a small conventional area (40 -200 m, aerodrome, railway, see Fig. 4), and is accelerated (with 0.7 -3 g) in air up to a speed of 270–300 m/s (Mach number 0.9) by the drive station on the distance 1–6 km. The glider flies (for distance 30–70 km, Figs. 2), gradually loses speed and increases attack angle of wing. When the speed drops so it is close to landing speed, the glider lands.

For take-off, instead of an aerodrome, a short (40-200 m) railway may be employed (Fig.4). The glider can be started from special bogie (trolley) up to speed 50 m/s, take off and then it is accelerated in air up to a speed of 250-300 m/s (Mach number 0.9) by the cable. After acceleration the cable is disconnected and glider free flights.

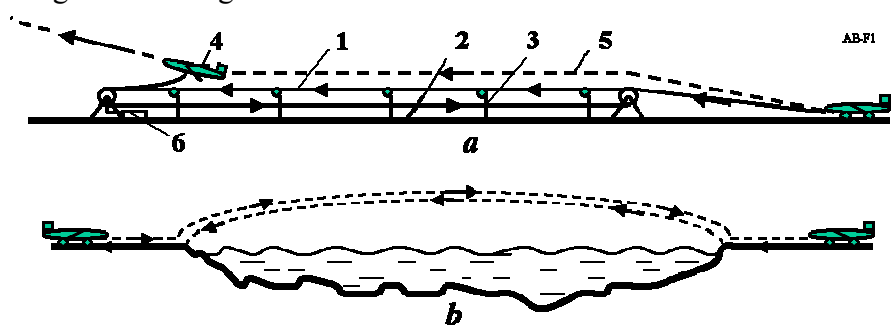

**Fig. 1.** JSAMT. (*a*) Terminal. Drive system for acceleration of kinetic glider (acceleration station); (*b*) Aerial Mega-Tramway across water. Notation: 1 – closed-loop cable; 2 – Earth's ground; 3 – support column with roller; 4 – flight glider; 5 – trajectory of flight vehicle; 6 – drive station.

The flight data are significantly improved if the kinetic aviation vehicle (glider) has variable wing area or variable swept wings[11]. The flight altitude does not influence its range because the energy spent in climbing will be returned in gliding. The Java-Sumatra Aerial Mega-Tramway offers the following noteworthy advantages:

(a) the load capability of kinetic aircraft—standard shipping containers—increases as a factor of two since the winged container has no fuel or engine;
(b) the kinetic aircraft, the attachable/detachable wings, is significantly less expensive than convention aircraft with a pressurized cabin (no engines and expensive navigation devices);
(c) the ground-based propulsion engine can operate on the cheapest available fuel;
(d) the maximum flight period is just a few minutes at most.

## Theory of kinetic vehicles and a general estimation of flight data

1. **The maximum range**, $R$, of kinetic air vehicles is obtained from the kinetic energy of theoretical mechanics. It is equals

$$d\left(\frac{mV^2}{2}\right) = \frac{mg}{K} dR, \quad R \approx \frac{K}{2g}\left(V_1^2 - V_0^2\right), \tag{1}$$

where $R$ is range [m]; $K$ is the average aerodynamic efficiency ($K$ = 10–20 for subsonic air vehicles and $K$ = 5–8 for supersonic air vehicles. For example: the subsonic Boeing-747 has maximum $K$ = 16, the recently retired supersonic "Concorde" has maximum $K$ = 7,5, supersonic aircraft XB-70 and YF-12 have $K$ = 7, and Boeing 2707-300 has $K$ = 7.8); $g$ = 9.81 m/s$^2$ is gravity; $V_1$ is initial (after acceleration) speed [m/s]; $V_0 < V_1$ is final (near landing) speed [m/s] ($V_0$ = 50–60 m/s); $V$ is variable speed, $V_0 < V < V_1$ [m/s], $mg/K = D$ is air drag [N]; $m$ is vehicle mass [kg]. Last equation in (1) is obtained from the first equation using integration. Results of our computations for subsonic ($V < 300$ m/s, $M < 0.9$, $M$ is Mach number) and supersonic ($M$ = 1 - 3) vehicles are presented in Figs. 2 and 3. The range of a subsonic vehicle is 45–90 km for $V_1$ = 300 m/s; the range of a supersonic vehicle can reach up 400 km for $V_1$ = 1000 m/s.

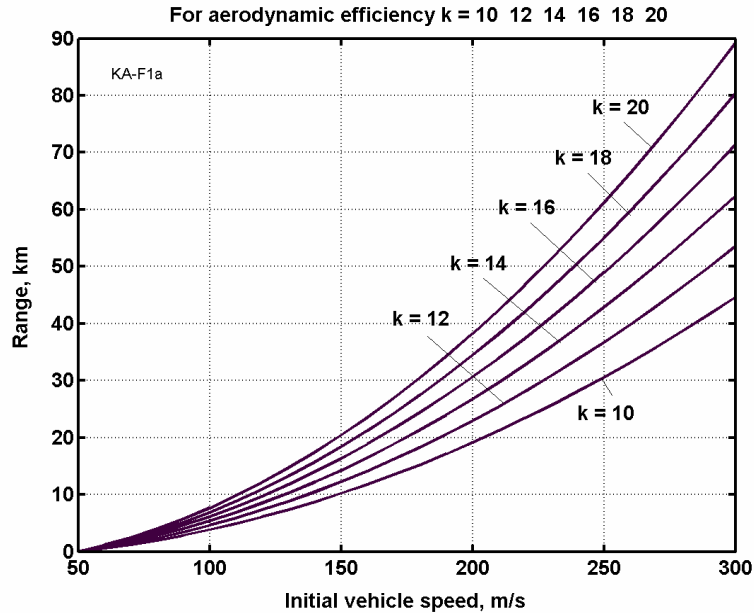

**Fig. 2.** Range of the subsonic kinetic glider versus initial speed for different aerodynamic efficiency $K$ = 10 12 14 16 18 20.

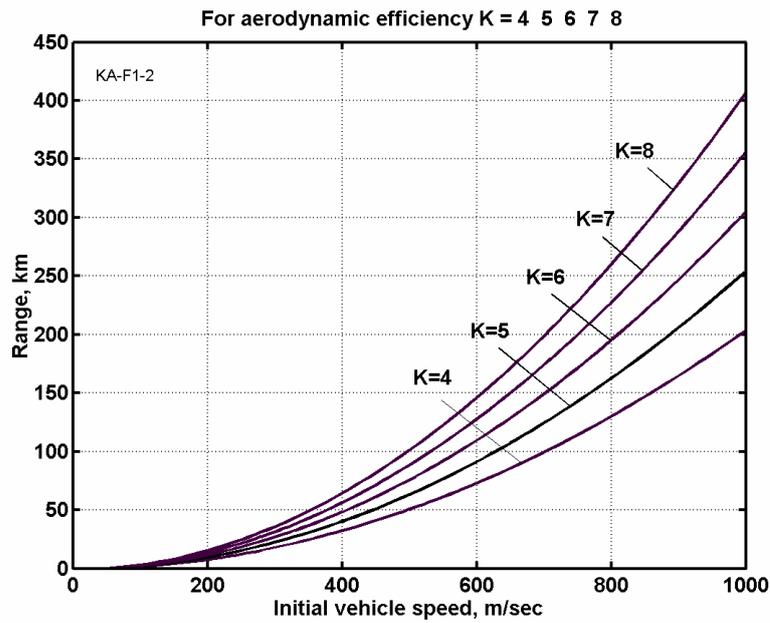

**Fig. 3.** Range of the supersonic kinetic glider versus initial speed for different aerodynamic efficiency $K = 4\ 5\ 6\ 7\ 8$.

**2. Maximum acceleration** distance can be calculated using the equation

$$S = \frac{V_1^2}{2gn},\qquad(2)$$

where $n$ is overload, $g$. Computation results for both subsonic and supersonic modern aircraft are presented in Fig. 4.

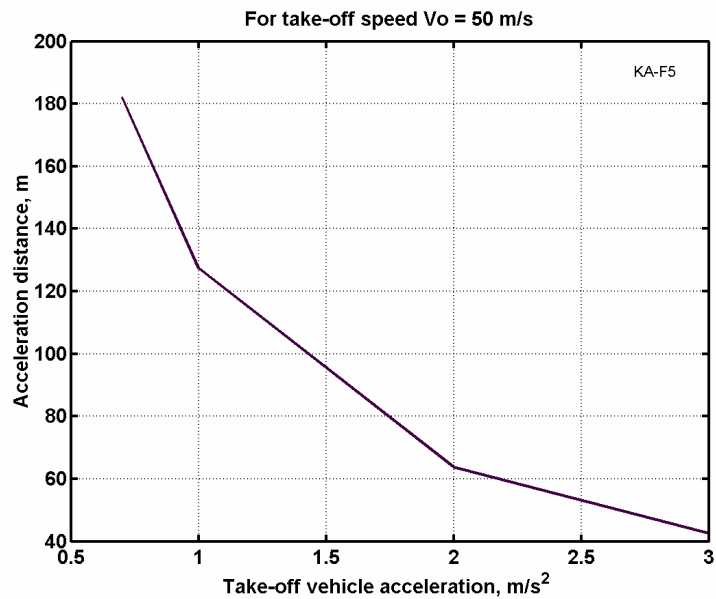

**Fig.4.** Aerodrome (railway) acceleration distance via a take-off acceleration of glider. Acceleration (3g) distance is 1500 m for a speed of 300 m/s for the subsonic vehicle and 17 km ($n = 3$) for a speed of 1 km/s for the supersonic vehicle (Figs. 5, 6). That is not aerodrome length—that is acceleration in air.

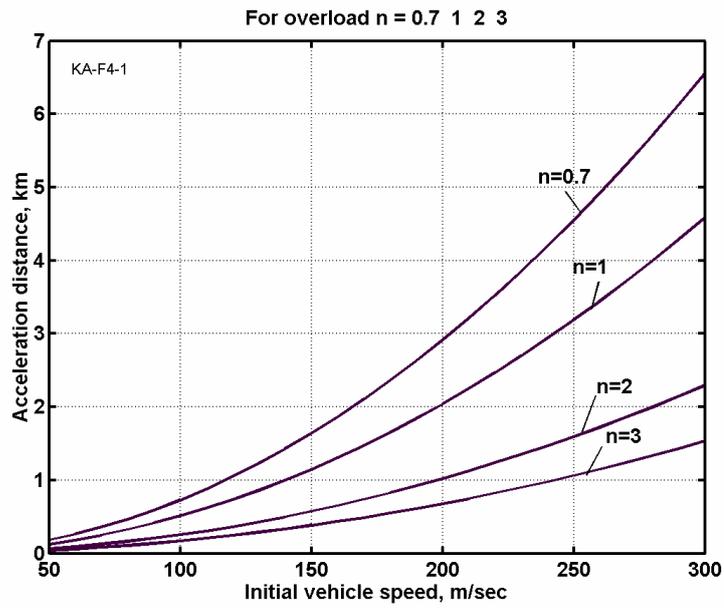

**Fig. 5**. Full acceleration distance (air acceleration is included) of subsonic kinetic glider versus an initial speed and different horizontal overloads.

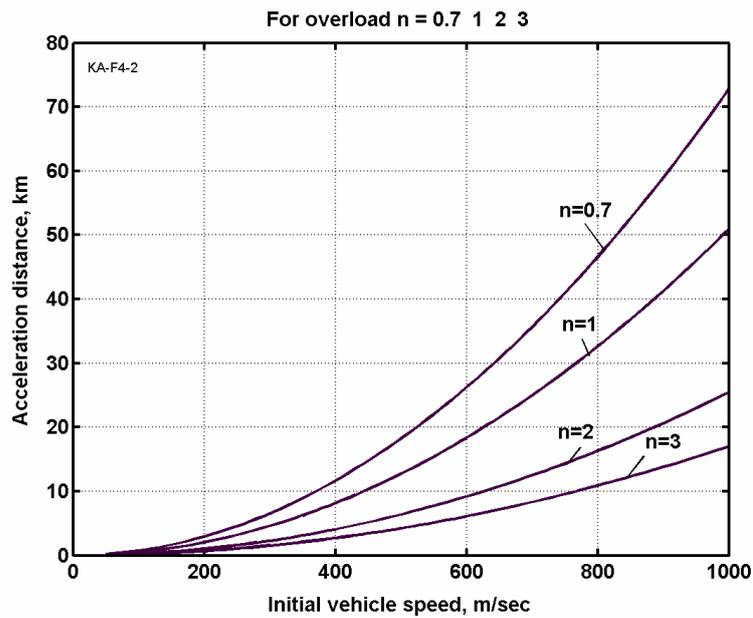

**Fig. 6**. Full acceleration distance (air acceleration is included) of supersonic kinetic glider versus an initial speed and different horizontal overloads.

3. **Average speed** and flight time are

$$V_a = \frac{V_1 + V_0}{2}, \quad T = \frac{R}{V_a}. \tag{3}$$

4. The trajectory of **horizontal turn** can be found from the following differential equations

$$\dot{V} = -\frac{gn}{K}, \quad \dot{\varphi} = \frac{L_1}{mV} = \frac{g\sqrt{n^2-1}}{V}, \quad \dot{x} = V\cos\varphi, \quad \dot{y} = V\sin\varphi, \quad \text{or}$$

$$V = V_1 - \frac{gn}{K}t > V_0, \quad \varphi = -\frac{K\sqrt{n^2-1}}{n}\ln\left(1 - \frac{gn}{K}t\right), \quad \dot{x} = V\cos\varphi, \quad \dot{y} = V\sin\varphi,$$

(4)

where $L_1$ is the projection of the vehicle lift force to a horizontal plane (vertical overload is 1); $t$ is time [seconds]; $\varphi$ is turn angle [rad].

Results of computations for different overloads are presented in Fig. 8. They show that the vehicle can turn back and return to its original aerodrome.

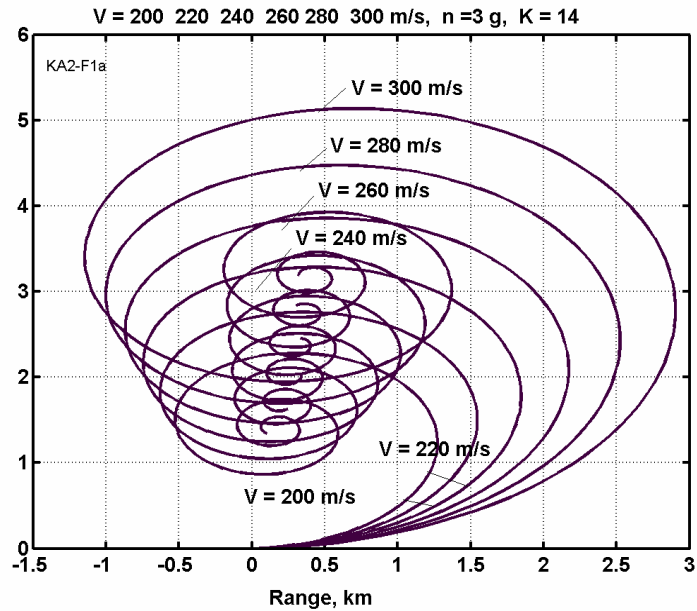

**Fig. 7.** Horizontal deviation versus range of the subsonic kinetic vehicle for initial speed $V = 200\ 220\ 240\ 260\ 280\ 300$ m/s, horizontal overload $n = 3$g, aerodynamic efficiency $K = 14$.

## 5. JSAMT Macroproject

Assume the mass of the flight vehicle is $m = 15$ tons (100 passengers and 4 members of crew); the acceleration is $a = 3g$ (this acceleration is acceptable for untrained people). The range is approximately $L = 35$ km (Sunda Strait, between islands Java and Sumatra as well as Palk Strait between India and Sri Lanka)(see Fig. 3 or calculate using Eq. (1) for a final acceleration speed of 250 m/s and $K = 14$, range is 43 km). The needed acceleration distance is $S = 1000$ m (Fig. 4). The starting area is only 40 m (Fig. 3). The time of horizontal acceleration is $t = (2S/a)^{0.5} = 8.2$ seconds. The flight time $t = 2 \times L/(V_1+V_0) = 467$ s = 7,8 minutes. Assuming it uses an artificial cheap fiber ($\sigma = 600$ kg/mm$^2$) widely produced by current industry, and a safety tensile strength of the drive vehicle cable is $\sigma = 180$ kg/mm$^2$ (the safety factor is $600/180 = 3.33$), density $\gamma = 1800$ kg/m$^3$ ($K_1 = 180/1800 = 0.1$). Then the cross-section area of the vehicle cable around the vehicle will be $S_1 = 3m/\sigma = 250$ mm$^2$, and the cable diameter is $d = 18$ mm. The mass of the cable is $M = S_1\gamma L_d = 450$ kg. Here $L_d = 1000$ m is the maximum length of the drive cable.

The energy required for acceleration of the aircraft and the cable is $E = mV^2/2$. This is about $E = 47$ Mega Joules (1 Mega Joule $= 10^6$ J) if $V = 250$ m/s. The drag of the aircraft and cable is about $D = 3$ 8tons, which means $E = DL = 3 \cdot 10^4 \cdot 1{,}000 = 39$ Mega Joules. If the launches are made launches are made every 0.1 hours and the ground-based engines must have a total power of about $P = E/t = 30 \cdot 10^6/6/60 = 83$ kW. If the engine efficiency is $\eta = 0.3$ the fuel consumption will be $F = E/\varepsilon/\eta = 39 \cdot 10^6/\varepsilon/0.3 = 3.1$ kg per flight. Here $\varepsilon = 42 \cdot 10^6$ [J/kg] is the energy capability of diesel fuel. This means that 0.031 kg of fuel is used for each passenger.

If tensile strength is $\sigma = 180$ kg/mm$^2$ = $1.8 \cdot 10^9$ N/m$^2$, $\gamma = 1800$ kg/m$^3$, then the total weight of the flywheels (as storage energy) will be about $M_w = 2E\gamma/\sigma = 2 \cdot 47 \cdot 10^6 \cdot 1800/1.8 \cdot 10^9 = 94$ kg.

Assume a cost of twenty million 2006 USA dollars for the installation, a lifetime of 20 years, and an annual maintenance cost of one million dollars. If 100 passengers are launched on every flight, there are 10 flights every hour for 350 days a year and the load coefficient is 0.75, then $N = 2 \times 100 \times 10 \times 24 \times 350 \times 0.75 = 12{,}600{,}000$ passengers will be launched per year. The launch cost per passenger is \$2,000,000/12,600,000 = \$0.16 plus fuel cost. If 0.031 kg of fuel is used for 1 passenger and the liquid fuel price is \$0.5 per kg, then the cost is \$0.016/person for liquid fuel. The total production cost will be about \$0.18/person. If each ticket costs but one USA dollar then the profit could be 10.3 millions dollars annually. The efficiency will be improved when the glider can take 200 and more passengers. For this short distance five (four working + one in reserve) gliders are enough. Fuel prices change with time, but in any case the cost of delivery will be sometimes less than delivery by conventional aircraft. The delivery time is less than any city transportation (bus, railway or subway)!

## 6. OTHER MACROPROJECT FACTORS

The public image of the JSAMT is at least as important as the macroproject's management during its physical creation[12]. Considering that the JSAMT is a novel freight and passenger moving innovation, some serious research must be undertaken in order to artfully craft a favorable public image of JSAMT that will enlighten and attract the public in the Republic of Indonesia. For example, many Indonesians will find computer simulations of a trip across the Sunda Strait aboard the winged containers too much "zoomscape", perhaps even frightening. JSAMT will be a pioneering new technology that must be presented locally both as a patented invention familiar enough not to alarm people and innovative enough so as not to seem too unfamiliar! One of the world's tallest office buildings, the Petronas Towers in Kuala Lumpur, has excited the public of Malaysia since 1996. So, public relations image managers will need to tune their public regional and global presentations to investors, government authorities and future JSAMT users carefully, taking into account all likely negative reactions.

The development of spacious well-lit and secure container storage yards at both ends of the JSAMT will require cooperation at many levels of government. There will be no need for transit sheds on Java or Sumatra since the standard shipping container is weather-tight and functions as a mobile warehouse. Refrigerated containers will require a reliable source of electricity. We view JSAMT as a unified/unitary transportation kinetic aviation system cooperating with railroad operators and truckers that has the potential to provide fixed linkages amongst, at least, the main islands of the Republic of Indonesia. JSAMT could reduce ship traffic in congested sea-lanes (Malacca Strait, Lomboc Strait and Sunda Strait) and help shippers avoid pirates plaguing slow-moving high-value cargo vessels. Once the JSAMT has been perfected, it may become adaptable to other Earth-surface sites such as the Bering Strait or Gibraltar Strait. At Palk Strait, the Aerial Mega-Tramway could lessen the ship traffic environmental impacts on Palk Bay.
   The other projects are in [13-18].